\documentclass[preprint,showpacs,preprintnumbers,amsmath,amssymb]{revtex4}

\usepackage[dvips]{graphicx}
\usepackage{dcolumn}
\usepackage{bm}

  \newcommand{\nn}{\nonumber}
  \newcommand{\sura}{\hspace{-1.5mm}\!/}
  
\begin{document}


\title{Jet substructures of boosted polarized top quarks}

\author{Yoshio Kitadono$^1$}
 \email{kitadono@phys.sinica.edu.tw}

\author{Hsiang-nan Li$^{1,2,3}$}
 \email{hnli@phys.sinica.edu.tw}
 \affiliation{$^1$Institute of Physics, Academia Sinica, Taipei 115, Taiwan, Republic of China}
 \affiliation{$^2$Department of Physics, National Cheng-Kung University,
 Tainan, Taiwan 701, Republic of China}
 \affiliation{$^3$Department of Physics, National Tsing-Hua University,
 Hsin-Chu, Taiwan 300, Republic of China}


\date{\today}

\begin{abstract}
We study jet substructures of a boosted polarized top quark, which
undergoes the semileptonic decay $t\to b\ell\nu$, in the perturbative QCD
framework. The jet mass distribution (energy profile) is factorized into the
convolution of a hard top-quark decay kernel with the bottom-quark jet
function (jet energy function). Computing the hard kernel to leading order
in QCD and inputting the latter functions from the resummation formalism,
we observe that the jet mass distribution is not sensitive to the helicity of
the top quark, but the energy profile is: energy is accumulated faster within
a left-handed top jet than within a right-handed one, a feature related to the
$V-A$ structure of weak interaction. It is pointed out that the energy
profile is a simple and useful jet observable for helicity discrimination of a
boosted top quark, which helps identification of physics beyond the Standard
Model at the Large Hadron Collider. The extension of our analysis to other jet
substructures, including those associated with a hadronically
decaying polarized top quark, is proposed.
\end{abstract}



\maketitle

\section{Introduction}

The precise theoretical and experimental investigation of top-quark
properties \cite{top.review1, top.review2,top.review3} is
crucial not only for understanding the electroweak dynamics in
the Standard Model, but also for exploring new physics beyond the Standard
Model. Especially, information of the top-quark polarization can reveal
the chiral structure of new physics. To translate the chiral couplings of
a top quark to new physics into observable polarization signals, the top
quark must be sufficiently boosted, as chirality is equivalent to helicity
in the massless limit. A top quark may be produced with large boost
at the Large Hadron Collider (LHC) in the future 14 TeV run, for example,
directly through the quark and antiquark annihilation, or indirectly through
the decay of a new massive particle. The final states of a boosted
top quark then become collimated and form a single jet. The polarization of
a top quark, if produced at rest, is determined by measuring the angular
distribution of decay products (see \cite{jet.func.spin.power,BGM14} and
references therein). Such measurement is not feasible even for the
final-state lepton with the largest spin analyzing power, as a
top quark is highly boosted.

Fixed-order calculations and soft-gluon resummations associated with the
boosted top quark production have been performed
in the Standard Model \cite{boost.top.SM1, boost.top.SM2,boost.top.SM3,boost.top.SM4}.
Exploration of new physics beyond the Standard Model by means of boosted top quarks
was studied intensively in \cite{boost.top.BSM1, boost.top.BSM2,boost.top.BSM3,
boost.top.BSM4, boost.top.BSM5, boost.top.BSM6, boost.top.BSM7, boost.top.BSM8,Ghosh:2013qga,
boost.top.BSM9, boost.top.BSM10}. Various strategies for tagging a boosted top
quark were proposed in \cite{boost.top.tagg1, boost.top.tagg2, boost.top.tagg3,
boost.top.tagg4, boost.top.tagg5, boost.top.tagg6, boost.top.tagg7, boost.top.tagg8},
and experimental searches have been conducted by ATLAS and CMS recently
\cite{boost.top.exp1, boost.top.exp2}. The relation between the polarization of
a boosted top quark and the energy fraction distribution of a particular subjet
was discussed in \cite{boost.top.pol} using
Monte Carlo generators. Given an algorithm for the subjet selection,
different energy fraction distributions for the left- and right-handed
top jets have been noticed. It implies that jet substructures can serve as
observables for distinguishing the helicity of a boosted polarized top quark.
In this paper we shall demonstrate in the perturbative QCD (pQCD) framework that the
energy profile of a top jet is a simple and useful substructure for this
purpose without requiring decomposition of subjets and algorithms for subject
selection as in \cite{boost.top.pol}, $b$-tagging, $W$-reconstruction or
measurement of missing momentum.

We start with the jet function of a polarized leptonic top, which
undergoes the $t\to b\ell\nu$ decay as an example. In pQCD this function
is factorized into the convolution of a hard top-quark decay kernel with the
bottom-quark jet function. The latter can be well
approximated by the light-quark jet function \cite{ALP09} derived in the
QCD resummation formalism \cite{energyprofile}, as the jet energy is high
enough. Evaluating the hard kernel to leading order (LO) in QCD, we obtain
the dependencies of the left- and right-handed top-quark jet functions
on the top-jet momentum and cone radius. It is found that the top-quark jet
function is not sensitive to the helicity as expected, since its peak
position is basically determined by the top-quark mass.
The similarity of the jet functions implies the similarity of the mass
distributions of the left- and right-handed top jets. Hence, boosted top
candidates can be identified by means of the jet mass measurement, together with
other promising techniques available in the literature. Certainly, the mass
measurement demands reconstruction of a missing neutrino momentum, which
we do not intend to explore further. The essence is
that additional information on the internal structure of a top jet
is required in order to distinguish its helicity.

After practicing the factorization of the top-quark jet function,
we extend it to the top-jet energy profile \cite{energyprofile.definition}
\begin{eqnarray}
\Psi(r) &=&
\frac{1}{N_{J_t}}\sum_{J_t}\frac{\sum_{r_i<r, i\in J_t}P_{T_i}}
{\sum_{r_i<R_t, i\in J_t}P_{T_i}},\label{profile}
\end{eqnarray}
where $N_{J_t}$ is the number of top jets with cone radius $R_t$, $r\le R_t$ is the radius
of a test cone centered around the top-jet axis, $P_{T_i}$ is the transverse momentum
carried by particle $i$ in the top jet $J_t$. The lepton energy is not included in
$\Psi(r)$, and the neutrino energy, as a missing momentum, does not contribute either.
This observable is then expressed as the convolution of a hard top-quark decay kernel with
the bottom-quark jet energy function. Inputting the light-quark jet energy function
derived also from the QCD resummation \cite{energyprofile}, we predict the energy
profiles of the left- and right-handed top jets. It turns out that the energy profile
is sensitive to the helicity: energy is accumulated faster within a
left-handed top jet than within a right-handed one, a feature related to
the $V-A$ structure of weak interaction. The dependencies of the energy profiles
of the left- and right-handed top jets on the top-jet momentum and cone
radius are presented for future experimental confrontation.

Our work does not only represent an application of pQCD to the study of jet
substructures of a boosted weakly decaying massive particle, but also highlights
the energy profile as a simple and useful observable for distinguishing the helicity
of a boosted top quark. The pQCD factorization formulas for the jet function and the
jet energy function of a polarized leptonic top are constructed in Sec.~\ref{formalism}.
The sensitivity of the top-jet mass distribution and energy profile
to the helicity is examined numerically in Sec.~\ref{massdist}.
Section~\ref{conclusion} is the conclusion.

\section{Formalism \label{formalism}}

\begin{figure}
  \begin{center}
    \def\SCALEa{0.7}
    \def\SCALEb{0.7}
    \def\OFFSET{5pt}
    \begin{tabular}{cc}
      \includegraphics[scale=\SCALEa]{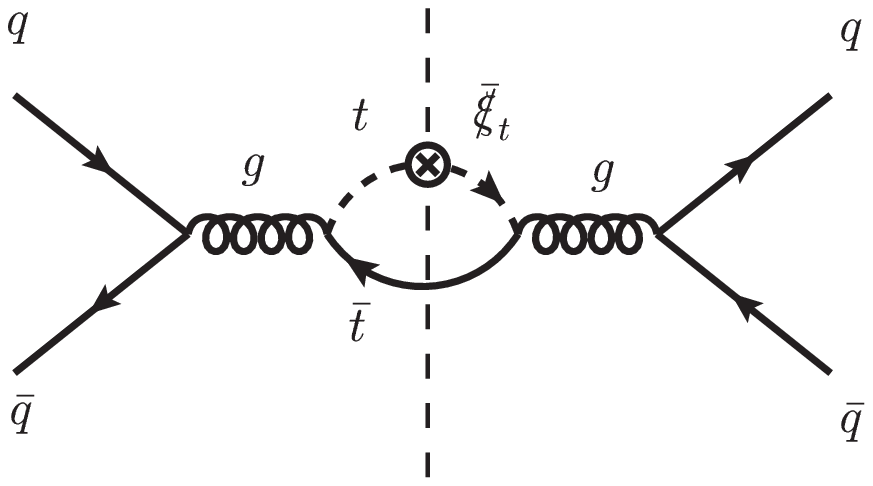} &
      \includegraphics[scale=\SCALEb]{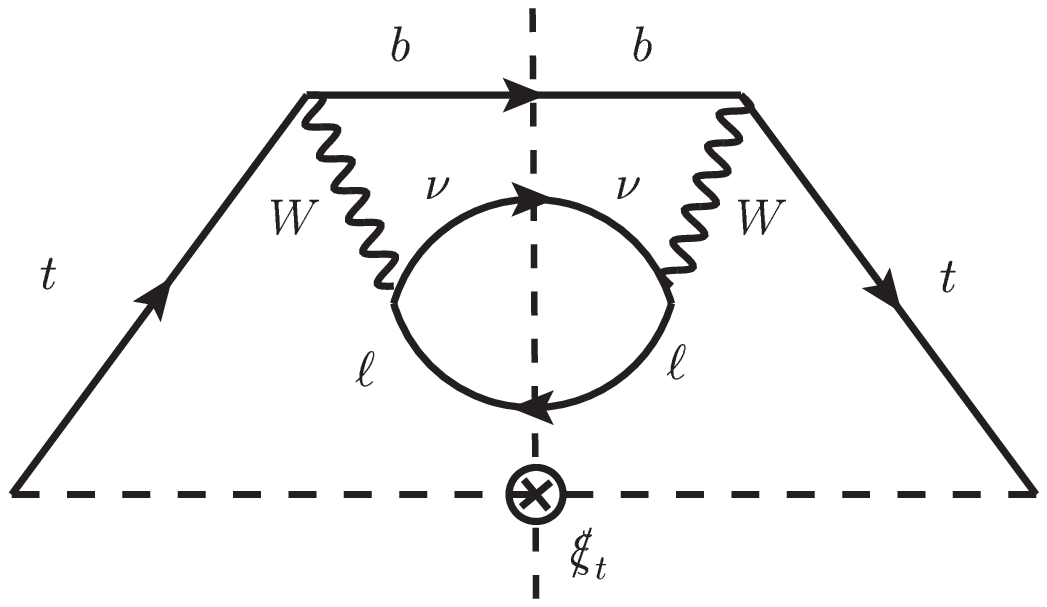} \\
      \hspace{\OFFSET} (a)
    & \hspace{\OFFSET} (b)
    \end{tabular}
    \caption{LO diagrams for (a) the production and (b) the semi-leptonic decay of a top quark.
    The dashed lines label the fermion flows, and the symbols $\otimes$ represent the insertions
    of the gamma matrices $\bar{\xi}\sura_{t}$ and $\xi\sura_{t}$ from the Fierz identity.}
    \label{fig.qqtt}
  \end{center}
\end{figure}

We consider only the annihilation $q\bar{q} \to t\bar{t}$ as the production
process for simplicity in order to demonstrate the construction of the
top-quark jet function $J_{t}$. Though the top-quark pair production is
largely dominated by the gluon-fusion channel at the LHC, the quark annihilation
channel is more important for the production of boosted top quarks,
which requires partons with large momentum fractions. Besides, our analysis
is not affected by the specific top-pair production channel as demonstrated
in the factorization procedure below. The factorization at LO in QCD is trivial, which
relies only on the insertion of the Fierz identity to break the fermion flow,
\begin{eqnarray}
 {\bf 1}_{ij}{\bf 1}_{lk}
&=& \frac{1}{4} {\bf 1}_{ik} {\bf 1}_{lj}
     + \frac{1}{4} (\gamma^{5})_{ik} (\gamma^{5})_{lj}
     + \frac{1}{4} (\gamma_{\alpha})_{ik} (\gamma^{\alpha})_{lj}
     + \frac{1}{4} (\gamma^{5}\gamma_{\alpha})_{ik}
                   (\gamma^{\alpha}\gamma^{5})_{lj}
     + \frac{1}{8} (\sigma_{\alpha\beta})_{ik}
                   (\sigma^{\alpha\beta})_{lj},\nn\\\label{fie}
\end{eqnarray}
with the identity matrix ${\bf 1}$, and $i$, $j$, $k$, and $l$ being the
Dirac indices.  We introduce the light-like vectors \cite{ALP09}
\begin{eqnarray}
 \xi^{\alpha}_{t} \equiv \frac{1}{\sqrt{2}}(1, -\vec{n}_{t}),
\hspace{2cm}
 \bar{\xi}^{\alpha}_{t} \equiv \frac{1}{\sqrt{2}}(1, \vec{n}_{t}),
\end{eqnarray}
to pick up the leading term in Eq.~(\ref{fie}). The unit vector $\vec{n}_t$
denotes the direction of the top-quark momentum, which
can be determined by measuring the hadronic anti-top in the $t\bar t$ system.
At higher orders, QCD radiations from the final states, that are collinear to the
top quark, are grouped into $J_{t}$ straightforwardly. The collinear initial-state
radiations are collected by the Wilson lines in the direction of $\xi_{t}$
\cite{energyprofile}, which are associated with the definition of $J_{t}$.

The corresponding LO matrix element squared $ \Big| \overline{\mathcal{M}} \Big|^2$
is then factorized, up to power corrections of $O(m_t/E_t)$, $m_t$ ($E_t$) being the
top-quark mass (energy), into two pieces, i.e., the production part
$ \Big| \overline{\mathcal{M}}_{pro} \Big|^2 $ shown in Fig.~\ref{fig.qqtt}(a) and
the decay part $ \Big| \overline{\mathcal{M}}_{decay} \Big|^2 $ for a leptonic top
in Fig.~\ref{fig.qqtt}(b),
\begin{eqnarray}
 \Big| \overline{\mathcal{M}} \Big|^2
&=&
 \Big| \overline{\mathcal{M}}_{pro} \Big|^2
 \Big| \overline{\mathcal{M}}_{decay} \Big|^2
\left[1 + O\left(\frac{m_t}{E_t}\right)\right],\nn\\
\Big| \overline{\mathcal{M}}_{pro} \Big|^2
&=& \frac{1}{4} \frac{g^4_s T_RC_F}{4N_cs^2}
    \mbox{tr}\left(\gamma^{\mu} k\sura \gamma^{\rho} \bar{k}\sura
	     \right)
    \mbox{tr}\left[\gamma_{\mu} (k\sura_{\bar{t}} - m_t)
	           \gamma_{\rho} \bar{\xi}\sura_{t}
	     \right], \nn\\
 \Big| \overline{\mathcal{M}}_{decay} \Big|^2
&=& C_{decay}
   \mbox{tr}
    \left( \gamma^{\gamma}P_{L}k\sura_{\nu}\gamma^{\sigma}P_{L}
	            k\sura_{\ell} \right)
    \mbox{tr}
     \left[\gamma_{\sigma}P_{L}(k\sura_t + m_t) \xi\sura_{t}
	            (k\sura_{{t}} + m_t) \gamma_{\gamma}P_{L}
		    (k\sura_b + m_b)
     \right],\label{fact}
\end{eqnarray}
where the factor $C_{decay}$ collects the $W$-boson and top-quark
propagators
\begin{eqnarray}
 C_{decay} = \frac{g^4}{4}|V_{tb}|^2
    \frac{1}{(q^2_W - m^2_{W})^2 + m^2_{W}\Gamma^2_{W}}
    \frac{1}{(k^2_t - m^2_{t})^2 + m^2_{t}\Gamma^2_{t}}.
\end{eqnarray}
In the above expressions, $g_s$ is the QCD coupling, $N_c=3$, $T_R=1/2$ and $C_F=4/3$ are the color
factors, $\sqrt{s}$ is the center-of-mass energy of the $t\bar{t}$ system,
$g$ is the weak coupling, $|V_{tb}|$ is the
Cabibbo-Kobayashi-Maskawa matrix element,  $k$, $\bar{k}$,
$k_t$, $k_{\bar{t}}$, $k_b$, $k_{\ell}$, $k_{\nu}$, and $q_W$ are the
momenta of the $q$ quark, $\bar{q}$ quark, $t$ quark, $\bar{t}$ quark,
$b$ quark, lepton, neutrino, and $W$ boson, respectively,
$m_{W}$ and $m_{b}$ are the masses of the $W$ boson
and $b$ quark, respectively, $\Gamma_{W}$ and $\Gamma_{t}$
are the decay widths of the $W$ boson and top quark, respectively, and
$P_{L}=(1-\gamma^5)/2$ is the chiral projection matrix.
The top-quark (anti-top-quark) line has been assigned to the decay
(production) piece in the factorization.

Next we decompose the decay piece for the unpolarized top quark
according to the spin states $s_t$ by introducing the projectors
\begin{eqnarray}
 w_t = \frac{1}{2}(1 + \gamma^5 s\sura_t), \hspace{2cm}
 \bar{w}_t = \frac{1}{2}(1 - \gamma^5 s\sura_t).
\end{eqnarray}
Equation~(\ref{fact}) becomes
\begin{eqnarray}
 \Big| \overline{\mathcal{M}}_{decay} \Big|^2
&=&    \Big| \overline{\mathcal{M}}^{s_t}_{decay} \Big|^2
     + \Big| \overline{\mathcal{M}}^{\bar{s}_t}_{decay} \Big|^2, \nn\\
\Big| \overline{\mathcal{M}}^{s_t}_{decay} \Big|^2
&=& 2C_{decay}(\xi_t \cdot k_t)
   \mbox{tr}
    \left( \gamma^{\gamma}P_{L}k\sura_{\nu}\gamma^{\sigma}P_{L}
	            k\sura_{\ell} \right)
   \mbox{tr}
    \left[\gamma_{\sigma}P_{L} (k\sura_t + m_t) w_{t}
          \gamma_{\gamma}P_{L} (k\sura_b + m_b)
     \right] \nn\\
&{}& + O(k^2_t - m^2_t), \nn\\
\Big| \overline{\mathcal{M}}^{\bar{s}_t}_{decay} \Big|^2
&=& \Big| \overline{\mathcal{M}}^{s_t}_{decay} \Big|^2_{w_t \to \bar{w}_t},
\end{eqnarray}
where the neglected terms are proportional to the virtuality $k^2_t - m^2_t$ of the
top quark.

\begin{figure}
  \begin{center}
    \def\SCALE{0.6}
      \includegraphics[scale=\SCALE]{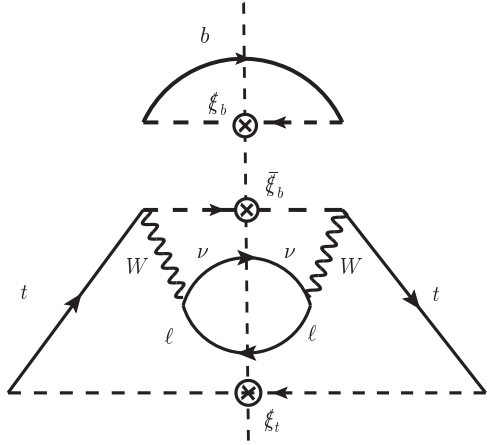}
    \caption{LO factorization of the bottom-quark jet function from the
    polarized leptonic top. The symbols $\otimes$ represent the insertions
    of the gamma matrices $\xi\sura_b$ and
    $\bar{\xi}\sura_b$ from the Fierz identity.}
    \label{fig.bfactorization}
  \end{center}
\end{figure}

Below we focus on the first decay piece
$\Big| \overline{\mathcal{M}}^{s_t}_{decay} \Big|^2$, which contributes to
the LO polarized top-quark jet function $J_{t}^{s_t(0)}(m^2_{J_t},E_{J_t},R_t)$.
After including QCD radiations, the top-jet mass $m_{J_t}$ (energy $E_{J_t}$)
may differ from the top-quark mass $m_t$ (energy $E_t$), i.e., the top-jet
momentum $k_{J_t}$ may differ from the top-quark momentum $k_t$. $J_{t}^{s_t(0)}$
can be factorized, as depicted in Fig.~\ref{fig.bfactorization},
by inserting the Fierz identity and by introducing the light-like vectors
\begin{eqnarray}
 \xi^{\alpha}_{b} \equiv \frac{1}{\sqrt{2}}(1, -\vec{n}_{b}),
\hspace{2cm}
 \bar{\xi}^{\alpha}_{b} \equiv \frac{1}{\sqrt{2}}(1, \vec{n}_{b}),
\end{eqnarray}
with the unit vector $\vec{n}_b$ being along the bottom-quark momentum.
The identity for defining the bottom-quark jet function with the
jet momentum $k_{J_b}$ \cite{jet.func.spin.power},
\begin{eqnarray}
 1 &=& \int dm^2_{J_b}dE_{J_b} d^2\vec{n}_{J_b} \delta(m^2_{J_b} - m^2_{b})
 \delta(E_{J_b} - E_{b}) \delta^2(\vec{n}_{J_b} - \vec{n}_b),
\end{eqnarray}
is also inserted, where $m_{J_b}$, $E_{J_b}$, and $\vec{n}_{J_b}$ are the
invariant mass, the energy, and the direction of the bottom jet,
respectively, and $E_{b}$ is the energy of the bottom quark. For
factorization at higher orders in QCD, the bottom-quark mass $m_b$
(energy $E_b$, direction $\vec{n}_{b}$) is replaced by the invariant
mass (total energy, direction of total momentum) of all the partons
in the bottom jet.

The LO polarized top-quark jet function is then written as
\begin{eqnarray}
 J_{t}^{s_t(0)}(m^2_{J_t},E_{J_t},R_t)
 &=& \frac{(2\pi)^3}{4\sqrt{2}E_{J_t}}\int dm^2_{J_b} dE_{J_b} d^2\vec{n}_{J_b}
 \frac{d^4k_{\ell}}{(2\pi)^3}\delta_+(k^2_{\ell})
 d^4k_{\nu}\delta_+(k^2_{\nu})
 \nn\\
&{}&\times H^{(0)}(k_{J_t},k_{J_b},k_{\ell}) J_{b}^{(0)}(m^2_{J_b}, E_{J_b}, R_b)
\delta^4(k_{J_t} - k_{J_b} - k_{\ell} - k_{\nu}),
\label{fact0}
\end{eqnarray}
with the LO top-quark decay kernel $H^{(0)}$ and
the LO bottom-quark jet function $J_{b}^{(0)}$
\begin{eqnarray}
H^{(0)}(k_{J_t},k_{J_b},k_{\ell})
&=& 4C_{decay}A_{b}^{-1}(\xi_t \cdot k_{J_t})
 (k_{\nu}\cdot \bar{\xi}_b)\left[ (k_{\ell}\cdot k_{J_t}) - m_t (k_{\ell}\cdot s_t)\right], \nn\\
 J_{b}^{(0)}(m^2_{J_b}, E_{J_b}, R_b)
 &=& A_b \int \frac{d^4k_b}{(2\pi)^3}\delta_{+}(k^2_b - m^2_b)
 \delta(m^2_{J_b} - m^2_b)\delta(E_{J_b} - E_b)\delta^2(\vec{n}_{J_b} - \vec{n}_b)
\nn\\
 &{}&
\times \mbox{tr}\left[(k\sura_b + m_b)\xi\sura_b \right],
\end{eqnarray}
$R_b$ being the bottom-jet cone radius.
The normalization constant $A_b=(2\pi)^3/(2\sqrt{2}E^2_{J_b})$
is absorbed into $J_{b}^{(0)}$, such that $J_{b}^{(0)}=\delta(m^2_{J_b} - m^2_b)$
holds after the integration over $k_b$ is worked out.
To arrive at Eq.~(\ref{fact0}), contributions down by powers of $m_{J_b}/E_{J_b}$ have
been ignored. The above factorization can be extended to all orders in QCD following
the procedures outlined in \cite{jet.func.spin.power}, and we replace
$J^{s_t(0)}_{t}$ ($J_b^{(0)}$) by its all-order definition
$J^{s_t}_{t}$ ($J_b$). The choice $R_b=R_t$ is made, so that the bottom-quark
jet function absorbs not only collinear radiations, but also soft radiations
in the top jet. The remaining infrared finite radiations go into the hard
top-quark decay kernel, which sticks to the LO expression below.

The lepton kinematic variables are first integrated out
in the rest frame of the top quark, leading to
\begin{eqnarray}
 J^{s_t}_{t}(m^2_{J_t},\bar E_{J_t},\bar R_t)
 &=&f_{t}(z_{J_t})\int dz_{J_b}d\bar x_{J_b}d\cos\bar\theta_{J_b}\nn\\
 &{}& \times
\left[  F_{a}(z_{J_t}, \bar x_{J_b}, z_{J_b}) + |\vec{s}_t|
F_{b}(\bar x_{J_b}, z_{J_b}) \cos\bar\theta_{J_b}
\right] J_{b}(m^2_{J_b}, \bar E_{J_b}, \bar R_t),\label{eq.Jtst}
\end{eqnarray}
where $\bar E_{J_t}=m_{J_t}$, $\bar E_{J_b}$ is the bottom-jet energy in the
rest frame, the dimensionless parameters $z_{J_t}$,
$\bar x_{J_b}$, and $z_{J_b}$ are defined as
\begin{eqnarray}
 z_{J_t} &=& \frac{m^2_{J_t}}{m^2_t}, \hspace{1cm}
 \bar x_{J_b} = \frac{2\bar E_{J_b}}{m_{J_t}}, \hspace{1cm}
 z_{J_b} = \frac{m^2_{J_b}}{m^2_{J_t}},
\end{eqnarray}
$\bar\theta_{J_b}$ is the polar angle of the bottom-jet momentum relative
to the top spin $\vec{s}_t$ (see Fig.~3 in Ref.~\cite{jet.func.spin.power}), and
$\bar R_t$ is the upper bound of $\bar\theta_{J_b}$ in the rest frame.
The hard functions $F_{a}$ and $F_{b}$ are given by
\begin{eqnarray}
 F_{a}(z_{J_t}, x_{J_b}, z_{J_b})
 &=& \sqrt{z_{J_t}}x_{J_b}\beta_{J_b} f_{W}(x_{J_b}, z_{J_b})
\left( - \frac{1}{3}x^2_{J_b} + \frac{1 + z_{J_b}}{2}x_{J_b} - \frac{2}{3}z_{J_b}
\right), \nn\\
F_{b}(x_{J_b}, z_{J_b})
 &=&  f_{W}(x_{J_b}, z_{J_b})
\left[ - \frac{1}{3}x^3_{J_b} + \frac{1 + 3z_{J_b}}{6}x^2_{J_b} + \frac{4}{3}z_{J_b}x_{J_b}- \frac{2}{3}z_{J_b}(1 + 3z_{J_b})
\right],
\end{eqnarray}
with $\beta_{J_b}=\sqrt{1- m^2_{J_b}/E^2_{J_b}}$,
which coincide with Eq.~(26) in the study of the spin analyzing power
for a polarized top quark at rest \cite{jet.func.spin.power}, as the on-shell
condition $z_{J_t}=1$ is taken. That is, Eq.~(\ref{eq.Jtst}) can be regarded
as a formula of the spin analyzing power for an off-shell top quark.
The functions $f_t$ and $f_{W}$
\begin{eqnarray}
f_{t}(z_{J_t}) &=&
\frac{1}{16\pi} |V_{tb}|^2m^4_{W}G^2_{F}z^{3/2}_{J_t}
\frac{1}{(1 - z_{J_t})^2 + \eta^2_t}, \nn\\
f_{W}(x_{J_b}, z_{J_b})
&=& \frac{1}{(1 + z_{J_b}- x_{J_b}- \xi)^2 + (\xi\eta)^2},
\end{eqnarray}
arise from the top-quark and $W$-boson propagators, respectively,
with the Fermi constant $G_{F}$, and the dimensionless mass ratios
\begin{eqnarray}
\eta_t = \frac{\Gamma_t}{m_t}, \hspace{1cm}
\xi = \frac{m^2_{W}}{m^2_{J_t}}, \hspace{1cm}
\eta = \frac{\Gamma_{W}}{m_W}.
\end{eqnarray}

The top-quark spin is chosen in the positive $z$-axis, $s^{\mu}_{t} = (0,0,0,1)$.
We then boost the top-quark rest frame with a velocity
parameter $v_t$ along the negative $z$-axis to make a right-handed top quark
($v_t>0$) or a left-handed top quark ($v_t<0$) \cite{Shelton}, with the absolute
value $|v_t|=\sqrt{1-m_{J_t}^2/E_{J_t}^2}$.
The jet energies $E_{J_t}$ and $E_{J_b}$,
and the polar angle $\theta_{J_b}$ of the bottom jet in the
boosted frame are related to those in the rest frame of the top quark
via the Lorentz transformation
\begin{eqnarray}
 \gamma_t\bar E_{J_t} &=& E_{J_t} ,\nn\\
 \bar x_{J_b} &=& 2\gamma^2_t {x}_{J_b} (1 - v_t \cos\theta_{J_b}),\nn\\
\cos\bar\theta_{J_b} &=& \frac{- v_t + \cos{\theta}_{J_b}}{1 - v_t \cos{\theta}_{J_b}},
\label{lor}
\end{eqnarray}
with the gamma factor $\gamma_t$, and the bottom-jet energy
fraction ${x}_{J_b} = {E}_{J_b}/E_{J_t}$.
We have neglected the bottom-jet mass compared to the top-jet
mass in Eq.~(\ref{lor}), since this approximation holds well and renders our
formalism simpler.

We then derive the right-handed top-quark jet function
\begin{eqnarray}
 J^{R}_{t}(m^2_{J_t}, E_{J_t}, R_t)
 &=&f_{t}(z_{J_t})\int_{0}^{1} dz_{J_b} \int_{\cos R_t}^{1}d\cos{\theta}_{J_b}
 \int_{0}^{J/(4\gamma^2_t)} d{x}_{J_b}\nn\\
 &{}& \times
J\left[  F_{a}(z_{J_t}, \bar x_{J_b}, z_{J_b}) + |\vec{s}_t|
F_{b}(\bar x_{J_b}, z_{J_b}) \cos\bar\theta_{J_b}
\right] J_{b}(m^2_{J_b}, E_{J_b}, R_t),
\label{eq.mass.dist.master}
\end{eqnarray}
with the Jacobian $J=2/(1-v_t\cos{\theta}_{J_b})$ being from the Lorentz transformation.
Note that the variables $\bar x_{J_b}$ and $\cos\bar\theta_{J_b}$ are understood as
functions of ${x}_{J_b}$ and $\cos{\theta}_{J_b}$ through Eq.~(\ref{lor}).
Moreover, we have replaced the bottom-quark jet function
$J_{b}(m^2_{J_b}, \bar E_{J_b}, \bar R_t)$ in Eq.~(\ref{eq.Jtst}) by
$J_{b}(m^2_{J_b}, E_{J_b}, R_t)$ in Eq.~(\ref{eq.mass.dist.master}). The use of the
former corresponds to the sequence of performing the factorization of the bottom-quark jet
function in the top-quark rest frame, and then applying the boost, while the use of the
latter corresponds to the sequence of applying the boost first, and then performing
the factorization. These two sequences are equivalent, because the jet function
depends on the product $E_{J_b}R_b$, instead of on $E_{J_b}$ and $R_b$ separately, as
observed in \cite{energyprofile}. It can be shown by employing the Lorentz transformation
in Eq.~(\ref{lor}) that $E_{J_b}R_b$, which bears the meaning of the transverse momentum
relative to the bottom-quark jet axis, is boost-invariant in the small $R_b$ limit.
Note that the factorization presented in this work and the resummation performed
in \cite{energyprofile} hold up to power corrections of $R_b$. In other words,
the scaling of a jet function with the product of the jet energy and the jet
cone radius is closely related to the commutability of the boost and factorization
operations. The left-handed top-quark jet function $J_{t}^{L}$, as a consequence of
the Lorentz transformation of $J_{t}^{\bar{s}_t}$, is easily obtained
from $J_{t}^{R}$ by flipping the sign of the second term, namely,
$\cos\bar\theta_{J_b}$ in Eq.~(\ref{eq.mass.dist.master}).

According to Eq.~(\ref{profile}), we count the contribution from a final-state
parton to the top-quark jet energy function, if it is emitted into the test
cone of radius $r$. It is assumed that the charged lepton can be identified
and its contribution is excluded. The neutrino does not contribute either
due to missing momentum. The LO factorization is applied to the expression
$|\overline{\mathcal{M}}_{decay}|^2 E_{J_b}\theta(r-\theta_{J_b})$, which
can be interpreted as the expectation value of the bottom-jet energy within
the test cone. This step results in the LO bottom-quark jet energy function
$J^{E(0)}_b=E_{J_b}\delta(m^2_{J_b} - m^2_b)$. The factorization
of QCD radiations gives, with the top-jet mass
being integrated out, the right-handed top-quark jet energy function
\begin{eqnarray}
 J^{E,R}_{t}(E_{J_t},R_t,r)
 &=&\int_0^{\gamma^2} \frac{dz_{J_t}}{z_{J_t}} f_{t}(z_{J_t})
 \int_{\cos r}^{1}d\cos{\theta}_{J_b}
 \int_{0}^{J/(4\gamma^2_t)} d{x}_{J_b}\nn\\
 &{}& \times
J x_{J_b}^2\left[  F_{a}(z_{J_t}, \bar x_{J_b}, 0) +
F_{b}(\bar x_{J_b}, 0) \cos\bar\theta_{J_b}
\right] J^{E}_{b}(E_{J_b}, R_t,r),
	\label{eq.energypro.master}
\end{eqnarray}
as the convolution of the hard functions $F_a$ and $F_b$ with the bottom-quark jet
energy function $J_b^{E}(E_{J_b}, R_t,r)$, where we have set $|s_t|=1$.
The smaller cone of radius $r$ in the bottom jet
is centered around the bottom-jet axis. The additional factor $x_{J_b}^2$ in the second line
of the above expression arises from the different Mellin variables
$m_{J_t}^2/(E_{J_t} R_t)^2$ for the left-hand side and $m_{J_b}^2/(E_{J_b} R_t)^2$
for the right-hand side. The bottom-jet mass dependence has been dropped
in the hard functions due to its smallness compared to other mass scales.
Therefore, the bottom-jet mass can be integrated out trivially, leading to the
$N=1$ moment of the bottom-quark jet energy function $J_b^{E}(E_{J_b}, R_t,r)$ in
Eq.~(\ref{eq.energypro.master}). The left-handed top-quark jet energy function
$J^{E,L}_{t}$ is derived from $J_{t}^{E,R}$ by flipping the sign of
$\cos\bar\theta_{J_b}$.

The lower bound $\cos r$ of $\cos{\theta}_{J_b}$  in Eq.~(\ref{eq.energypro.master})
implies that the bottom jet does not contribute to the top-jet energy profile, as
the polar angle of the bottom-jet axis goes outside the test cone.
However, the smaller cone in the bottom jet still overlaps
with the test cone when $\theta_{J_b}$ is slightly greater than $r$, so
the contribution from the bottom jet does not vanish sharply. How to
count the partial contribution from the overlap region of the test cone
of the top jet and the smaller cone of the bottom jet depends on
a scheme, under which radiations in the former are factorized into
the latter. Our goal in this work is to demonstrate the difference
between the energy profiles of the left- and right-handed top jets.
Hence, we do not intend to explore the sophisticated issue related
to the factorization scheme, and assume vanishing of the bottom-jet
contribution as $\theta_{J_b}>r$. At last, the hard functions contain
an additional factor $1-v_t$ in the boosted frame actually, indicating 
that the decay of a highly-boosted right-handed top quark with 
$v_t \to 1$ is power suppressed as expected.
However, this factor cancels in the ratio in Eq.~(\ref{profile})
defined for the energy profile, and is not shown explicitly.


\section{NUMERICAL ANALYSIS\label{massdist}}

It is easy to obtain the polarized top-quark jet functions at LO in
QCD by substituting $\delta(m^2_{J_b}- m^2_{b})$ for the bottom-quark
jet function $J_{b}$ in Eq.~(\ref{eq.mass.dist.master}).
For the inclusion of QCD effects, we replace $J_b$ in
Eq.~(\ref{eq.mass.dist.master}) by the light-quark
jet function from the resummation formalism \cite{energyprofile}.
It is more convenient to employ the fitted expression in Eq.~(29) of
Ref.~\cite{jet.func.spin.power}.
The following parameters \cite{PDG} are adopted
\begin{eqnarray}
m_{b}&=&4.18~\mbox{GeV}, \hspace{1cm}
m_{W}=80.39~\mbox{GeV},\hspace{1cm}m_{t} =172.5~\mbox{GeV},\nn\\
G_{F}&=& 1.16637\times10^{-5}~\mbox{GeV}^{-2}, \hspace{1cm}
\Gamma_{W}= 2.09~\mbox{GeV},\nn\\
\Gamma^{\mbox{\tiny LO}}_{t}&=& 1.48~\mbox{GeV},
\hspace{1cm}\Gamma^{\mbox{\tiny NLO}}_{t}=1.33~\mbox{GeV},
\end{eqnarray}
where the top-quark decay width $\Gamma^{\mbox{\tiny LO}}_{t}$
is inserted into the function $f_t$ for the analysis at LO, and
the next-to-leading-order (NLO) one $\Gamma^{\mbox{\tiny NLO}}_{t}$
\cite{NLO.topwidth} is inserted for the analysis with resummation.

\begin{figure}
  \begin{center}
    \def\SCALE{0.6}
         \includegraphics[scale=\SCALE]{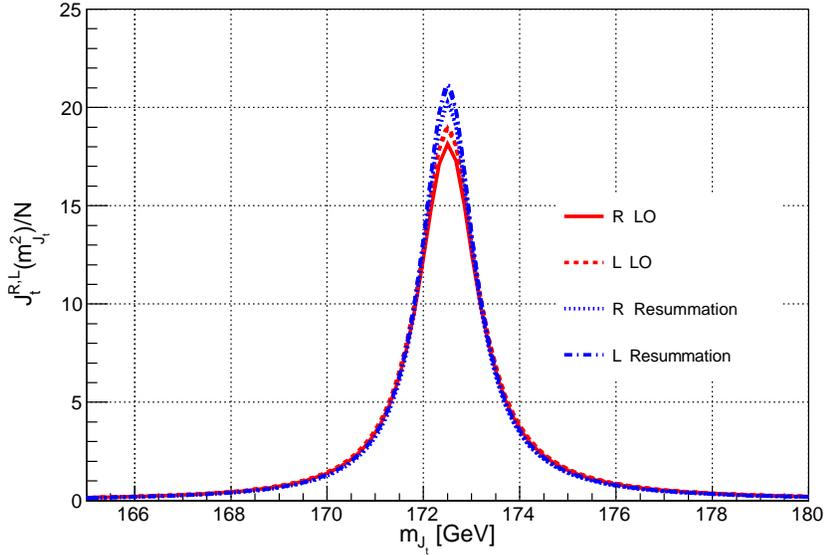}
    \caption{Comparison between the LO results and those including the resummation effect
    for the polarized top-quark jet functions with $E_{J_t}=1$ TeV and $R_t=0.7$.}
    \label{fig.mass.dist.QCD}
  \end{center}
\end{figure}

The jet mass dependence of the normalized top-quark jet functions
$J^{R,L}_{t}(m^2_{J_t})/N$, $N=\int dm^2_{J_t}[J_{t}^R(m^2_{J_t})+J_{t}^L(m^2_{J_t})]$,
for the top-jet energy $E_{J_t}=1$ TeV and cone radius $R_t=0.7$ are
displayed in Fig.~\ref{fig.mass.dist.QCD}. As mentioned before, the
bottom-quark jet function demands that the dominant contribution arises
from the region $m_{J_b} \sim O(m_b)$, which is much smaller than other
mass scales, like the top-quark and $W$-boson masses. If the
bottom-jet mass dependence is negligible in the hard kernel, the
integration of the bottom-quark jet function over the bottom-jet mass
can be performed trivially, giving a normalization constant. Therefore,
the LO curves are close to those with the resummation effect. It has been
numerically confirmed that the lower
LO values around the peak position are mainly attributed to the top decay
widths $\Gamma^{\mbox{\tiny LO}}_{t}>\Gamma^{\mbox{\tiny NLO}}_{t}$. It is also
observed that the left-handed jet function is roughly the same as the
the right-handed one, and slightly higher than the right-handed one at the peak
position. It implies that the mass distributions of the left- and right-handed top
jets, computed as the convolution of the parton-level production cross section with parton
distribution functions and the polarized top-quark jet functions, will be basically
identical. We have also investigated the behavior of the normalized top-quark jet
functions for $E_{J_t}=500$ GeV, 1 TeV, and 2 TeV and for $R_t=0.4$, 0.7, and 1.0,
which is similar to that shown in Fig.~\ref{fig.mass.dist.QCD}.
The difference between the left- and right-handed top-quark jet functions
decreases with $E_{J_t}$ as expected.

We then focus on the energy profiles of the polarized top jets.
For a top jet produced in $q\bar{q} \to t\bar{t}$ at central rapidity with
a fixed top-jet energy, its energy profile is simply related to the jet energy
function, because the production pieces, including the LO parton-level cross
section and parton distribution functions, cancel in the ratios
\begin{eqnarray}
R(r) &=& \frac{J_{t}^{E,R}(E_{J_t},R_t, r)}{J_{t}^{E,R}(E_{J_t},R_t,R_t)},\hspace{2cm}
L(r) = \frac{J_{t}^{E,L}(E_{J_t},R_t, r)}{ J_{t}^{E,L}(E_{J_t},R_t, R_t)}.
\end{eqnarray}
It is obvious that the above definitions for the right- and left-handed top-jet
energy profiles satisfy $R(r=0)= L(r=0)=0$ and $R(r=R_t)= L(r=R_t)=1$. We calculate
the LO jet energy function $J^{E,R}_{t}$ for the right-handed top quark by
substituting $J_b^{E(0)} = 1/E_{J_b}$ in the Mellin space for the bottom-quark
jet energy function in Eq.~(\ref{eq.energypro.master}). As verified above, the
bottom-jet mass dependence in the factorization formula for the top-quark jet function
is negligible. With this approximation, the analytical expressions for the LO
top-quark jet energy functions can be derived, which are presented in the appendix.

\begin{figure}
  \begin{center}
    \def\SCALE{0.5}
      \includegraphics[scale=\SCALE]{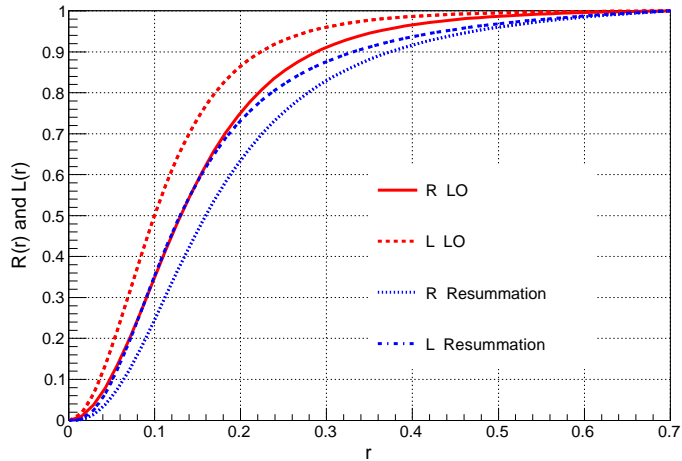}
    \caption{Comparison between the LO results and those including the resummation effect
    for the energy profiles of the polarized top jets with $E_{J_t}=1$ TeV and $R_t=0.7$.}
    \label{fig.energypro.QCD}
  \end{center}
\end{figure}

The QCD effects are taken into account by adopting the bottom-quark jet energy
function $J_{b}^E=(1/E_{J_b}) \exp[S_{q}({E}_{J_b}, {R}_{t}, r)]$,
which resums the double logarithm $\alpha_s\ln^2(r/R_t)$ to all orders
\cite{energyprofile}.
The Sudakov exponent $S_{q}({E}_{J_b}, {R}_{t}, r)$ is evaluated
with the LO QCD running coupling. In principle, the NLO correction to the initial
condition $1/E_{J_b}$ of the bottom-quark jet energy function should be included,
which is acquired by matching the Sudakov exponent to the complete
NLO contribution. However, this correction, without the large logarithm,
is expected to be less crucial. The LO results of the energy
profiles $R(r)$ and $L(r)$ and those from the QCD resummation for
$E_{J_t}=1$ TeV and $R_t=0.7$ are compared in Fig.~\ref{fig.energypro.QCD}: the
curves with the resummation are lower than the LO ones by 15\% at $r=0.2$,
by 30\% at $r=0.1$, and by almost 100\% as $r\to 0$. The QCD effects spread the
bottom-quark energy into soft gluons, such that the energy is accumulated more slowly

\begin{figure}
  \begin{center}
    \def\SCALE{0.5}
    \begin{tabular}{c}
      \includegraphics[scale=\SCALE]{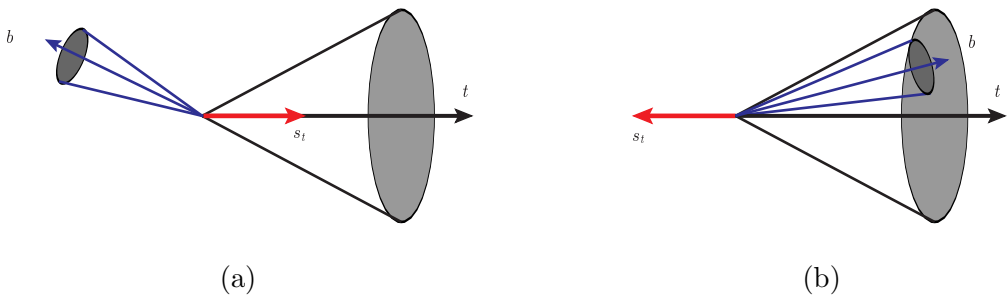}\\
     \hspace{0cm}  (a) \hspace{7cm} (b)
    \end{tabular}
    \caption{Pictorial explanation of the different energy profiles in the (a) right-handed and
    (b) left-handed top jets.  The configuration (b) shows a larger probability for the
    bottom jet to go inside the top-jet cone due to the $V-A$ structure $(\kappa_b < 0)$.}
    \label{fig.left.dominance}
  \end{center}
\end{figure}

It is also noticed that the jet energy is accumulated faster in the left-handed
top jet than in the right-handed one. The mechanism is the same as that
responsible for the higher mass distribution of the left-handed top jet,
which can be understood via the $V-A$ structure of weak interaction.
First, the spin analyzing power $\kappa_i$ for decay product $i$
of a polarized top quark in the rest frame is
defined via the angular distribution \cite{Spin.analyz1,Spin.analyz2}
\begin{eqnarray}
 \frac{1}{\Gamma}\frac{d\Gamma}{d\cos\theta_{i}}
  &=& \frac{1}{2}(1 + \kappa_i \cos\theta_{i}), \hspace{1cm}i=b,\ell,\nu,
\end{eqnarray}
where $\Gamma=\Gamma(t\to b\ell \nu)$ is the partial decay width, and $\theta_i$
is the polar angle of the decay product momentum relative to the top spin.
The spin analyzing powers for the bottom quark, the charged lepton, and the neutrino
were found to be $\kappa_b \simeq -0.4$, $\kappa_{\ell}\simeq +1.0$, and
$\kappa_{\nu}\simeq -0.3$ (see the Table 1 in \cite{Spin.analyz3} with the
correspondence between the $u$ quark and the neutrino, and between the $\bar{d}$
quark and the charged lepton). That is, the bottom quark and the neutrino tend to
be emitted in the direction opposite, i.e., anti-correlated to the top spin, while
the charged lepton tends to be correlated to the top spin.
According to the definition of the helicity, the bottom quark, which tends to be
emitted opposite to the top-quark momentum, has a higher chance to go outside
the cone of a right-handed top jet as depicted in
Fig.~{\ref{fig.left.dominance}}(a). Hence, the bottom jet gives less contribution to
the right-handed top-jet mass and energy profile. For a boosted left-handed top
quark, the favored configuration is the one with the bottom quark being emitted
along the top-quark momentum, so the bottom jet has a higher chance to go inside
the top-jet cone as shown in Fig.~{\ref{fig.left.dominance}}(b),
and contributes more to the jet mass and energy profile.
The difference between the energy fraction distributions of a particular subjet
for the left- and right-handed top jets observed in \cite{boost.top.pol}
is also attributed to the same mechanism.

\begin{figure}
  \begin{center}
    \def\SCALE{0.42}
    \def\OFFSET{10pt}
    \begin{tabular}{cc}
      \includegraphics[scale=\SCALE]{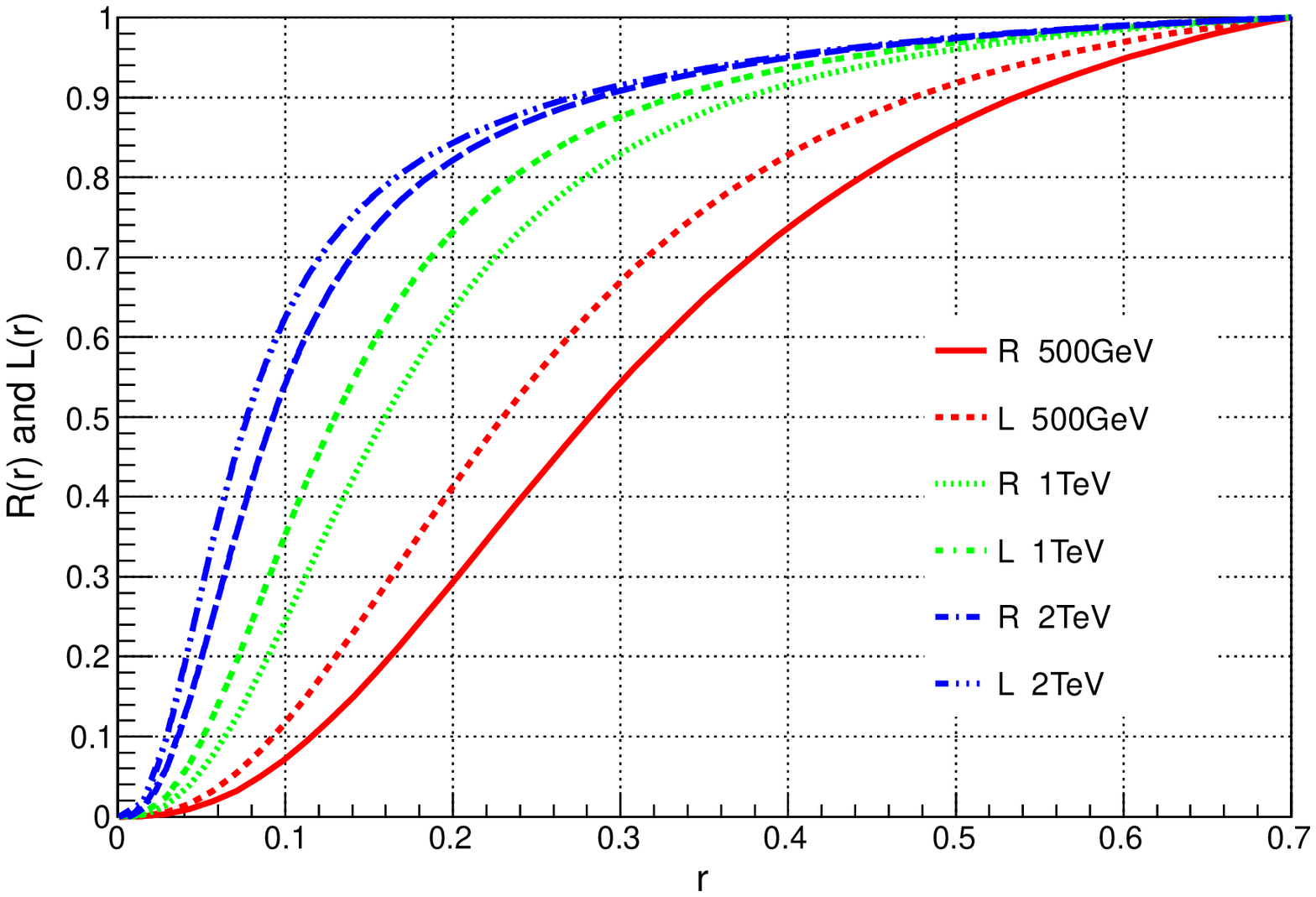} &
      \includegraphics[scale=\SCALE]{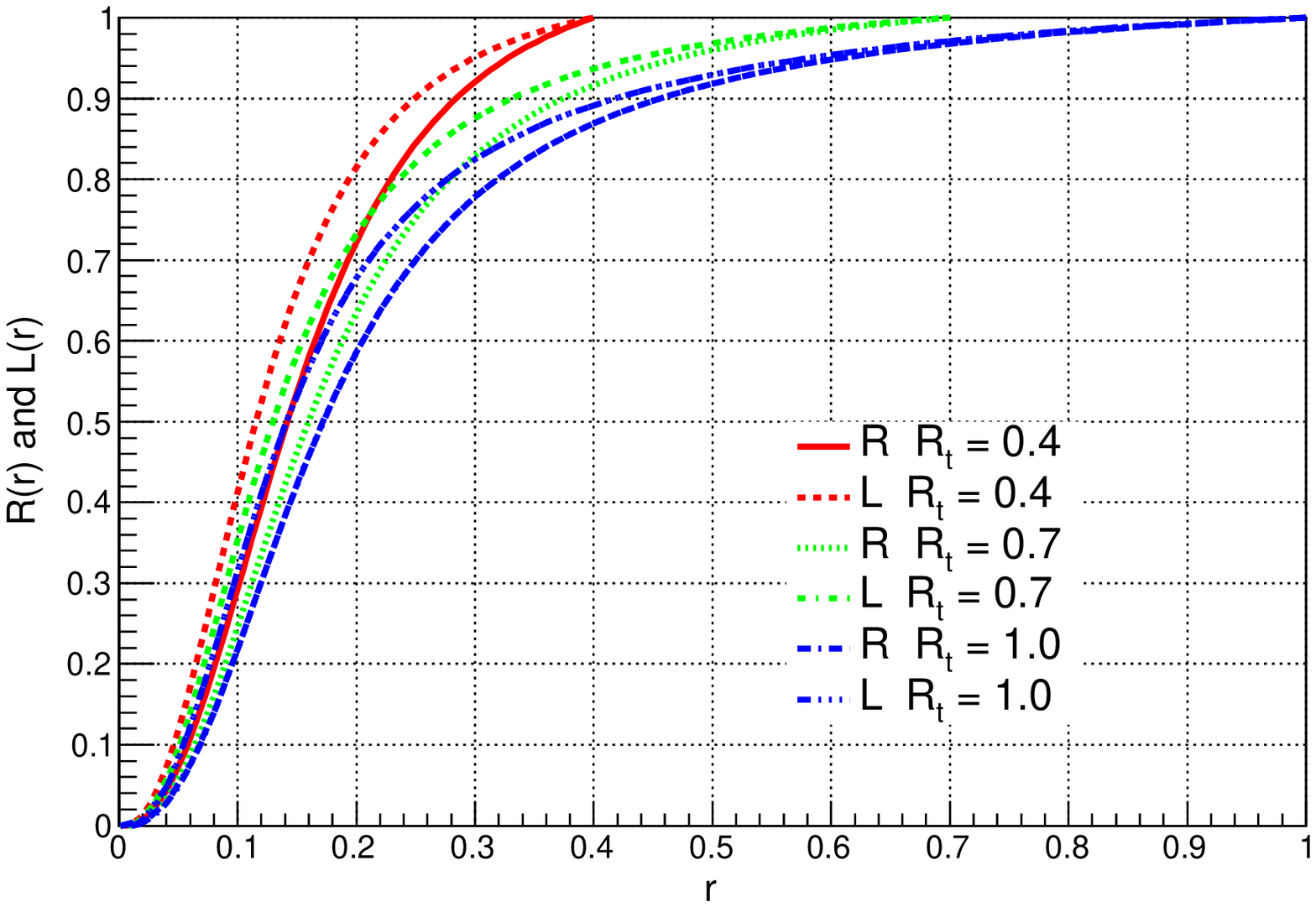}\\
       \hspace{\OFFSET} (a)
    & \hspace{\OFFSET} (b)
    \end{tabular}
    \caption{Dependencies of the top-jet energy profiles on (a) top-jet energy $E_{J_t}$
    for $R_t=0.7$ and (b) top-jet cone radius $R_{t}$ for $E_{J_t}=1$ TeV.}
    \label{fig.energypro.Ejt.Rjt}
  \end{center}
\end{figure}

The energy profiles $R(r)$ and $L(r)$ for various top-jet energies $E_{J_t}=500$ GeV,
1 TeV, and 2 TeV and cone radii $R_t=0.4$, 0.7, and 1.0 are presented in
Fig.~\ref{fig.energypro.Ejt.Rjt}. For a typical transverse momentum ($< 1$ TeV)
carried by a boosted top quark at the LHC, $R(r)$ and $L(r)$ can differ by about
30\% at small $r<0.2$ as shown in Fig.~\ref{fig.energypro.Ejt.Rjt}(a), which is significant
enough for experimental discrimination: To distinguish the curves of $R(r)$ and
$L(r)$ for $E_{J_t}=1$ TeV up to $5\sigma$ level (statistical error only),
thousands of boosted top quarks are needed. With the normalized distribution
$(1/\sigma_{t\bar{t}})d\sigma_{t\bar{t}}/dm_{t\bar{t}}\approx 10^{-2}$ at the top-pair
invariant mass $m_{t\bar{t}}\approx 2$ TeV \cite{Aad:2012hg}, and the total cross
section $\sigma_{t\bar{t}}\approx 170$ pb \cite{ATLAS}, we can have a sufficient
number of boosted top quarks in the central pseudorapidity interval $|\eta|<0.5$
and in the mass interval $1.5\; \mbox{TeV}<m_{t\bar{t}}< 2.5\;\mbox{TeV}$ for the integrated
luminosity about 10 fb$^{-1}$. As a top quark becomes highly boosted, the difference
decreases as expected, since the bottom quark becomes collimated with the top quark
more exactly. For example, the difference reduces to about few percent at $r=0.2$ as the
top-jet energy reaches $E_{J_t}=2$ TeV for $R_t=0.7$. Figure~\ref{fig.energypro.Ejt.Rjt}(b)
indicates that the difference between $R(r)$ and $L(r)$ is not sensitive to the cone
radius as $R_t>0.7$. For $R_t=0.4$, the difference drops to 10\% at $r=0.2$, and the
energy profiles do not exhibit saturation in the whole range of $r$.
It implies that this cone radius is too small to collect majority of the top-jet energy,
and a cone radius larger than $R_t=0.7$ is preferred for LHC measurements.

\section{Conclusion \label{conclusion}}
In this paper we have studied the helicity dependence of the mass distribution and
the energy profile as examples of jet substructures for a boosted polarized
top quark. The former (latter) is factorized into the convolution of the
hard top-quark decay kernel with the bottom-quark jet function (jet energy function).
It has been found that the QCD effects introduced by the bottom-quark jet function
on the mass distribution of a polarized top jet are minor.
Those introduced by the bottom-quark jet energy function
on the top-jet energy profile are more significant, which lower
the LO top-jet energy profile by 15\% at $r=0.2$ and by 30\% at $r=0.1$
for the top-jet energy $E_{J_t}=1$ TeV and cone radius $R_t=0.7$.
The reduction is similar for the right- and left-handed top jets,
since QCD interaction is vector-like, i.e., it is independent of the chirality
of the top quark.

Both the jet mass distribution and the jet energy profile of the left-handed
top jet are larger than those of the right-handed one for various top-jet
energies and cone radii. This observation is a consequence of the $V-A$ structure
of weak interaction, under which the favored direction of the emitted bottom quark
is opposite to the top spin. However, the mass distribution is not sensitive to
the helicity, but the energy profile is: energy is accumulated faster within the
left-handed top jet than within the right-handed one. The difference is about
30\% at small $r$ with a larger top-jet cone $R_t>0.7$ for typical boosted top
quarks at the LHC. The measurement of the
neutrino missing momentum, the $b$-tagging, and the $W$-reconstruction are not
required for this observable. That is, the energy profile is a simple and useful
jet substructure for helicity discrimination of a boosted top quark, which
can help identification of new physics beyond the Standard Model at the LHC.

We expect differences in other jet substructures of the left- and
right-handed top jets. A straightforward application of our formalism is to
study the leptonic energy distributions, whose difference between the left-
and right-handed top jets is also related to the $V-A$ structure of weak
interaction. An improved method to measure the polarization of a hadronic
top has been proposed in \cite{BT14}, in which the helicity
dependencies of top jet observables were explored. We will extend our
formalism to analysis of jet substructures
of a boosted hadronically decaying top quark.
In addition to the factorization of QCD radiations into
three light-quark jets, a soft function, which collects soft gluon exchanges
among the three final-state light quarks, needs to be introduced.
The behavior of this soft function can be extracted through the resummation
technique in the manner the same as for the light-particle jet functions
\cite{energyprofile}. The energy profile of a hadronically decaying top jet can
then be predicted.

\section*{Acknowledgment}
YK thanks H. Yokoya and members of physics departments at
Hiroshima university and Toyama university for useful discussions.
This work was supported in part by the National Science Council of
R.O.C. under Grant No. NSC-101-2112-M-001-006-MY3, and by the National
Center for Theoretical Sciences of R.O.C..

\appendix
\section{LO TOP JET ENERGY FUNCTIONS}

We compute the LO jet energy function $J^{E,R}_{t}$ for the right-handed
top quark by substituting $1/E_{J_b}$ for the bottom-quark jet energy function
$J_b^E$ in Eq.~(\ref{eq.energypro.master}). Neglecting the bottom-jet mass
dependence, we have
\begin{eqnarray}
 J^{E,R}_{t}( E_{J_t},R_t, r)
 &=& {f}^{E}_{t}
 \left[ \left( \frac{1}{2}I_{3}- \frac{1}{3}I_{4} \right)A( E_{J_t}, r)
 + \left( \frac{1}{6}I_{3}- \frac{1}{3}I_{4} \right)B(E_{J_t}, r)
\right],\label{je0}
\end{eqnarray}
with the constant
\begin{eqnarray}
{f}^{E}_{t} = \frac{1}{32}\frac{1+v^2_t}{\gamma_t^4 R_{t}E_{J_t}\eta_t }
 G^2_{F}m^4_{W} |V_{tb}|^2.
\end{eqnarray}
The functions
\begin{eqnarray}
 A(E_{J_t}, r)
&=& \frac{1}{2}\frac{(1 - \cos r)[2 - v_t(1+\cos r)] }{(1 - v_t)^2(1 - v_t \cos r)^2},\nn\\
 B(E_{J_t}, r)
&=& \frac{1}{6v^2_t}
\left[
\frac{(-1 + 2v_t)}{(1-v_t)^2}
+ \frac{1 - 3v_t \cos r + 2v^2_t }{(1 - v_t \cos r)^3}
\right],
\label{eq.AandB}
\end{eqnarray}
come from the integration over the polar angle of the bottom jet.
The functions
\begin{eqnarray}
I_3 &=&
\frac{5}{2} - 2\xi
+ \frac{1-\xi}{\xi\eta}\left[ (1-\xi)^2 - 3(\xi\eta)^2\right]
\left[ \tan^{-1}\left( \frac{1}{\eta}\right) -
\tan^{-1}\left( \frac{1}{\eta}\left( 1-\frac{1}{\xi}\right)\right)\right]
\nn\\
&{}& \hspace{0.1cm}
+ \frac{1}{2}\left[ 3(1-\xi)^2 - (\xi\eta)^2\right]
\ln\left[\frac{1+\eta^2}{(1-1/\xi)^2+\eta^2} \right],\nn\\
 I_{4} &=&
\frac{1}{3}
\left[  1 + 3(1-\xi) + 9(1-\xi)^2 - 3(\xi\eta)^2\right]
\nn\\
&{}& \hspace{0.1cm}
+ \frac{1}{\xi\eta}
\left[ (1-\xi)^4 - 6(1-\xi)^2(\xi\eta)^2 + (\xi\eta)^4\right]
\left[ \tan^{-1}\left( \frac{1}{\eta}\right) -
\tan^{-1}\left( \frac{1}{\eta}\left( 1-\frac{1}{\xi}\right)\right)\right]
\nn\\
&{}& \hspace{0.1cm}
+ 2(1-\xi)\left[ (1-\xi)^2 - (\xi\eta)^2\right]
\ln\left[\frac{1+\eta^2}{(1-1/\xi)^2+\eta^2} \right],
\label{eq.I1}
\end{eqnarray}
are results of the integration over the bottom-jet energy, where the subscripts
$3$ and $4$ in $I_3$ and $I_4$ refer to the powers of $\bar x_{J_b}$ in the
integrands. The narrow-width approximation has been applied to the top-quark propagator,
as performing the $z_{J_t}$ integration.
The LO jet energy function $J^{E,L}_{t}$ for the left-handed top quark is derived by flipping
the sign of the second term in Eq.~(\ref{je0}).


\end{document}